\documentclass[twocolumn,showpacs,aps,prd]{revtex4}
\usepackage{graphicx}
\usepackage{amsmath}
\usepackage{epsfig}

\newcommand{\BABARPubYear}    {05}
\newcommand{\BABARPubNumber}  {028}
\newcommand{\SLACPubNumber} {11332}

\RequirePackage{xspace}

\usepackage{relsize}
\def\babar{\mbox{\slshape B\kern-0.1em{\smaller A}\kern-0.1em
    B\kern-0.1em{\smaller A\kern-0.2em R}}}

\def\epem       {\ensuremath{e^+e^-}\xspace}

\def\ellell     {\ensuremath{\ell^+ \ell^-}\xspace}

\def\qqbar {\ensuremath{q\overline q}\xspace}

\def\pip   {\ensuremath{\pi^+}\xspace}
\def\pim   {\ensuremath{\pi^-}\xspace}
\def\pipi  {\ensuremath{\pi^+\pi^-}\xspace}
\def\pipm  {\ensuremath{\pi^\pm}\xspace}
\def\pimp  {\ensuremath{\pi^\mp}\xspace}

\def\Kbar  {\kern 0.2em\overline{\kern -0.2em K}{}\xspace}

\def\Kz    {\ensuremath{K^0}\xspace}
\def\Kzb   {\ensuremath{\Kbar^0}\xspace}
\def\KzKzb {\ensuremath{\Kz \kern -0.16em \Kzb}\xspace}
\def\Kp    {\ensuremath{K^+}\xspace}
\def\Km    {\ensuremath{K^-}\xspace}
\def\Kpm   {\ensuremath{K^\pm}\xspace}

\def\KpKm  {\ensuremath{\Kp \kern -0.16em \Km}\xspace}
\def\KS    {\ensuremath{K^0_{\scriptscriptstyle S}}\xspace}

\def\Dbar    {\kern 0.2em\overline{\kern -0.2em D}{}\xspace}

\def\Dz      {\ensuremath{D^0}\xspace}
\def\Dzb     {\ensuremath{\Dbar^0}\xspace}
\def\DzDzb   {\ensuremath{\Dz {\kern -0.16em \Dzb}}\xspace}
\def\Dp      {\ensuremath{D^+}\xspace}
\def\Dm      {\ensuremath{D^-}\xspace}

\def\DpDm    {\ensuremath{\Dp {\kern -0.16em \Dm}}\xspace}

\def\B       {\ensuremath{B}\xspace}
\def\Bbar    {\kern 0.18em\overline{\kern -0.18em B}{}\xspace}

\def\BB      {\ensuremath{B\Bbar}\xspace} 
\def\Bz      {\ensuremath{B^0}\xspace}
\def\Bzb     {\ensuremath{\Bbar^0}\xspace}
\def\BzBzb   {\ensuremath{\Bz {\kern -0.16em \Bzb}}\xspace}
\def\Bu      {\ensuremath{B^+}\xspace}
\def\Bub     {\ensuremath{B^-}\xspace}
\def\Bp      {\ensuremath{\Bu}\xspace}
\def\Bm      {\ensuremath{\Bub}\xspace}
\def\Bpm     {\ensuremath{B^\pm}\xspace}

\def\BpBm    {\ensuremath{\Bu {\kern -0.16em \Bub}}\xspace}

\def\BorBbar    {\kern 0.18em\optbar{\kern -0.18em B}{}\xspace}
\def\DorDbar    {\kern 0.18em\optbar{\kern -0.18em D}{}\xspace}
\def\KorKbar    {\kern 0.18em\optbar{\kern -0.18em K}{}\xspace}

\def\jpsi     {\ensuremath{{J\mskip -3mu/\mskip -2mu\psi\mskip 2mu}}\xspace}
\def\psitwos  {\ensuremath{\psi{(2S)}}\xspace}

\def\chiczero {\ensuremath{\chi_{c0}}\xspace}

\mathchardef\Upsilon="7107
\def\Y#1S{\ensuremath{\Upsilon{(#1S)}}\xspace}

\def\FourS {\Y4S}

\mathchardef\Deltares="7101
\mathchardef\Xi="7104
\mathchardef\Lambda="7103
\mathchardef\Sigma="7106
\mathchardef\Omega="710A

\def\Deltabar{\kern 0.25em\overline{\kern -0.25em \Deltares}{}\xspace}
\def\Lbar{\kern 0.2em\overline{\kern -0.2em\Lambda\kern 0.05em}\kern-0.05em{}\xspace}
\def\Sigbar{\kern 0.2em\overline{\kern -0.2em \Sigma}{}\xspace}
\def\Xibar{\kern 0.2em\overline{\kern -0.2em \Xi}{}\xspace}
\def\Obar{\kern 0.2em\overline{\kern -0.2em \Omega}{}\xspace}
\def\Nbar{\kern 0.2em\overline{\kern -0.2em N}{}\xspace}
\def\Xb{\kern 0.2em\overline{\kern -0.2em X}{}\xspace}

\def\mes        {\mbox{$m_{\rm ES}$}\xspace}

\def\DeltaE     {\mbox{$\Delta E$}\xspace}

\newcommand{\tev}{\ensuremath{\mathrm{\,Te\kern -0.1em V}}\xspace}
\newcommand{\gev}{\ensuremath{\mathrm{\,Ge\kern -0.1em V}}\xspace}
\newcommand{\mev}{\ensuremath{\mathrm{\,Me\kern -0.1em V}}\xspace}
\newcommand{\kev}{\ensuremath{\mathrm{\,ke\kern -0.1em V}}\xspace}
\newcommand{\ev}{\ensuremath{\mathrm{\,e\kern -0.1em V}}\xspace}
\newcommand{\gevc}{\ensuremath{{\mathrm{\,Ge\kern -0.1em V\!/}c}}\xspace}
\newcommand{\mevc}{\ensuremath{{\mathrm{\,Me\kern -0.1em V\!/}c}}\xspace}
\newcommand{\gevcc}{\ensuremath{{\mathrm{\,Ge\kern -0.1em V\!/}c^2}}\xspace}
\newcommand{\mevcc}{\ensuremath{{\mathrm{\,Me\kern -0.1em V\!/}c^2}}\xspace}

\def\cm   {\ensuremath{{\rm \,cm}}\xspace}

\def\invfb   {\ensuremath{\mbox{\,fb}^{-1}}\xspace}

\def\mus  {\ensuremath{\rm \,\mus}\xspace}

\def\mus        {\ensuremath{\,\mu{\rm s}}\xspace}    

\def\ra                 {\ensuremath{\rightarrow}\xspace}
\def\to                 {\ensuremath{\rightarrow}\xspace}

\def\pep2{PEP-II}

\def\gsim{{~\raise.15em\hbox{$>$}\kern-.85em
          \lower.35em\hbox{$\sim$}~}\xspace}
\def\lsim{{~\raise.15em\hbox{$<$}\kern-.85em
          \lower.35em\hbox{$\sim$}~}\xspace}

\def\CP                {\ensuremath{C\!P}\xspace}

\xspace

\newcommand{\jprlBase}       {Phys.\ Rev.\ Lett.\xspace}
\newcommand{\jprBase}        {Phys.\ Rev.\xspace}
\newcommand{\jplBase}        {Phys.\ Lett.\xspace}

\newcommand{\npBase}         {Nucl.\ Phys.\xspace}
\newcommand{\zpBase}         {Z.\ Phys.\xspace}

\newcommand{\ijmpa}     [1]  {{Int.\ J.\ Mod.\ Phys.\ {\bf A{\bf #1}}}}

\newcommand{\npb}       [1]  {\npBase\ B~{\bf #1}}

\newcommand{\plb}       [1]  {\jplBase\ B~{\bf #1}}

\newcommand{\jprl}      [1]  {\jprlBase\ {\bf #1}}
\newcommand{\pr}        [1]  {\jprBase\ {\bf #1}}
\newcommand{\jprd}      [1]  {\jprBase\ D~{\bf #1}}

\newcommand{\progtp}    [1]  {{Prog.\ Theor.\ Phys.\ {\bf #1}}}

\newcommand{\zp}        [1]  {\zpBase\ {\bf #1}}
\newcommand{\zpc}       [1]  {\zpBase\ C~{\bf #1}}

\def\jetset74   {\mbox{\tt Jetset \hspace{-0.5em}7.\hspace{-0.2em}4}\xspace}

\newcommand{\onreslumi}  {\mbox{210.3 \invfb}}
\newcommand{\offreslumi} {\mbox{21.6 \invfb}}
\newcommand{\bbpairs}    {\mbox{231.6 $\pm$ 2.6 million}}

\newcommand{\costtb}     {\mbox{$\cos{\theta_{T}}$}}
\newcommand{\abscosttb}  {\mbox{$\left|\cos{\theta_{T}}\right|$}}

\newcommand{\Btoppppos}         {\mbox{$\Bp \to \ppppos$}}
\newcommand{\Btopppneg}         {\mbox{$\Bm \to \pppneg$}}

\newcommand{\ppppos}            {\mbox{$\pip  \pip  \pim$}}

\newcommand{\pppneg}            {\mbox{$\pim  \pim  \pip$}}

\newcommand{\PPP}               {\mbox{$\pipm \pipm \pimp$}}
\newcommand{\KPP}               {\mbox{$\Kpm  \pimp \pipm$}}

\newcommand{\BtoPPP}            {\mbox{$\Bpm \to \PPP$}}
\newcommand{\BtoKPP}            {\mbox{$\Bpm \to \KPP$}}

\newcommand{\BtoKspi}           {\mbox{$\Bpm \to \KS\pipm$}}
\newcommand{\Kstopipi}          {\mbox{$\Ks \to \pip \pim$}}

\newcommand{\rhoz}          {\mbox{$\rho^0$}}

\newcommand{\rhoI}         {\mbox{$\rhoz(770)$}}
\newcommand{\rhoII}         {\mbox{$\rhoz(1450)$}}

\newcommand{\fz}            {\mbox{$f_0(980)$}}

\newcommand{\fzII}          {\mbox{$f_2(1270)$}}

\newcommand{\fzIII}         {\mbox{$f_0(1370)$}}

\newcommand{\Dzbpi}         {\mbox{$\Dzb \pip$}}

\newcommand{\BtoDzbpi}      {\mbox{$B^+ \to \Dzbpi$}}

\newcommand{\Dzbtopipi}      {\mbox{$\Dzb \to \pip\pim$}}
\newcommand{\DzbtoKpi}      {\mbox{$\Dzb \to \Kp\pim$}}

\newcommand{\JPsitoll}   {\mbox{$\jpsi \to \ellell$}}

\newcommand{\Psitoll}   {\mbox{$\psitwos \to \ellell$}}

\def\Y#1S{{\Upsilon\rm(#1S)}}
\def\Ks{{K^0_{\scriptscriptstyle S}}}

\def\ra{\rightarrow}
\def\etal{{\it et~al.,}}
\def\beq{\begin{equation}}
\def\eeq{\end{equation}}

\long\def\inst#1{\par\nobreak\kern 4pt\nobreak
    {\it #1}\par\vskip 10pt plus 3pt minus 3pt}

\begin{document}

\begin{flushleft}
\babar-PUB-\BABARPubYear/\BABARPubNumber \\
SLAC-PUB-\SLACPubNumber \\
\end{flushleft}

\title{\Large \bf \boldmath Amplitude Analysis of the Decay \BtoPPP}

%
\author{B.~Aubert}
\author{R.~Barate}
\author{D.~Boutigny}
\author{F.~Couderc}
\author{Y.~Karyotakis}
\author{J.~P.~Lees}
\author{V.~Poireau}
\author{V.~Tisserand}
\author{A.~Zghiche}
\affiliation{Laboratoire de Physique des Particules, F-74941 Annecy-le-Vieux, France }
\author{E.~Grauges}
\affiliation{IFAE, Universitat Autonoma de Barcelona, E-08193 Bellaterra, Barcelona, Spain }
% ...
\collaboration{The \babar\ Collaboration}
\noaffiliation

\date{August 2, 2005}

\begin{abstract}
We present a Dalitz-plot
analysis of charmless \Bpm\ decays to the final state
\PPP\ using 210 \invfb\ of data recorded by the \babar\ experiment at $\sqrt{s} = 10.58~\gev$. 
We measure the branching fractions
${\cal{B}}(\BtoPPP) = (16.2 \pm 1.2 \pm 0.9) \times 10^{-6}$ and
${\cal{B}}(B^{\pm} \ra \rhoI \pi^{\pm}) = (8.8 \pm 1.0 \pm 0.6^{+0.1}_{-0.7}) \times 10^{-6}$.
Measurements of branching fractions for the quasi-two-body decays
$B^{\pm} \ra \rhoII \pi^{\pm}$, $B^{\pm} \ra \fz \pi^{\pm}$ and
$B^{\pm} \ra \fzII \pi^{\pm}$ are also presented.
We observe no charge asymmetries for the above modes, and
there is no evidence for the decays $B^{\pm} \ra \chiczero \pi^{\pm}$,
$B^{\pm} \ra \fzIII \pi^{\pm}$ and $B^{\pm} \ra \sigma \pi^{\pm}$.

\end{abstract}


\maketitle

\setcounter{footnote}{0}

\section{Introduction}
\label{sec:Introduction}

The decay of $B$ mesons to a three-body charmless final state offers 
the possibility of investigating the properties of the weak
interaction and provides information on the complex quark couplings
described in the Cabibbo-Kobayashi-Maskawa (CKM) matrix 
elements~\cite{CKM:1973}, as well as on models of hadronic decays. 
Measurements of direct \CP-violating asymmetries and constraints 
on the magnitudes and the phases of the 
CKM matrix elements can be obtained from
individual decay channels in 
\BtoPPP~\cite{Gronau:1995,Blanco:1998,Blanco:2001,SnyderQuin:1993}, 
which are dominated by decays through intermediate resonances.
For example, the CKM angle~$\gamma$ can be extracted from
the interference between the decay $B^{\pm} \ra \chiczero \pi^{\pm}$, which has no
\CP-violating phase, and other modes such as $\rhoI \pi^{\pm}$ or $\fz \pi^{\pm}$.
Studies of these decays can also help to
clarify the nature of the resonances involved, not all of which are
well understood. Of particular interest
is whether the $\sigma$ resonance, which has been
observed in other experiments~\cite{Alde,E791,BESsigma}, is also 
present in \BtoPPP\ decays.
An analysis of the full three-body kinematic space is necessary to model the 
interference and extract branching fractions.

Observations of \B-meson decays to the \PPP\ three-body final states
have already been reported by the Belle and \babar\ collaborations
using a method that treats each intermediate decay
incoherently~\cite{Belle,Babar1}.
These studies have only observed $B^{\pm} \ra \rhoI \pi^{\pm}$, 
in which other possible resonance
contributions are treated as background. The first measurement of the
total branching fraction for \BtoPPP\ was found to be
$(11 \pm 4) \times 10^{-6}$~\cite{Babar2}.
Here, we present results from
a full amplitude analysis for \BtoPPP\ decay modes based on a
\onreslumi\ data sample containing \bbpairs\ \BB\ pairs collected with
the \babar\ detector~\cite{babardet} at the SLAC PEP-II
asymmetric-energy \epem\ storage ring~\cite{pep} operating at the
$\Upsilon(4S)$ resonance at a center-of-mass energy of
$\sqrt{s}=10.58$~\gev. An additional integrated luminosity of
\offreslumi\ was recorded $40$~\mev below this energy and
is used to study backgrounds. The charm
decay \BtoDzbpi, \DzbtoKpi~\cite{chargeConjugate} is used as a calibration channel 
as it presents a relatively high branching fraction.

\section{\boldmath The \babar\ Detector}
\label{sec:babar}

Details of the \babar\ detector are described elsewhere~\cite{babardet}. 
Charged particles are measured with
the combination of a silicon vertex tracker (SVT), 
which consists of five layers of
double-sided detectors, and a 40-layer central drift chamber (DCH) in a
1.5-T solenoidal magnetic field. This provides a transverse momentum
resolution for the combined tracking system of $\sigma_{p_T}/p_T =
0.0013p_T \oplus 0.0045$, where the sum is in quadrature and $p_T$ is
measured in \gevc.

Charged-particle identification is accomplished by
combining information on the specific ionization $(dE/dx)$ in the two
tracking devices and the angle of emission of Cherenkov radiation in an
internally reflecting ring-imaging Cherenkov detector (DIRC) covering
the central region. The $dE/dx$ resolution from the drift chamber is
typically about $7.5\,\%$ for pions. The Cherenkov angle resolution of
the DIRC is measured to be 2.4 mrad, for the quartz refractive index of 1.473,
which provides better than 3~$\sigma$
separation between charged kaons and pions over the full
kinematic range of this analysis.
The DIRC is surrounded by an electromagnetic calorimeter (EMC), comprising
6580 CsI(Tl) crystals, which is used to measure the energies
and angular positions of photons and electrons.
The EMC is used to veto electrons in this analysis.

\section{Event Selection and Reconstruction}
\label{sec:selection}

Hadronic events are selected based on track multiplicity and event topology.
Backgrounds from non-hadronic events are reduced by requiring the ratio of
Fox-Wolfram moments $H_2/H_0$~\cite{FoxWolfram} to be less than 0.98.
\B-meson candidates are reconstructed from events that have four or more 
charged tracks.
Each track is required to be well measured and originate from the beam spot.
They must have at least 12 hits in the DCH,
a minimum transverse momentum of 100 \mevc, and a distance of closest approach
to the beam spot of less than 1.5\cm\ in the transverse plane and
less than 10\cm\ along the beam axis. Charged tracks identified as electrons are rejected.
The \B-meson candidates are formed from three-charged-track
combinations and particle identification criteria are applied. 
The efficiency of selecting pions is approximately 95\,\%,
while the probability of misidentifying kaons as pions is 15\,\%.
The \B-meson candidates' energies and momenta are required to satisfy the kinematic
constraints detailed in Section~\ref{sec:finalsel}.

\section{Background Suppression}
\label{sec:background}

Backgrounds from $\epem \ra \qqbar$ are high and are suppressed by imposing
requirements on event-shape variables calculated in the $\FourS$ rest frame. 
The first discriminating variable is \costtb, the cosine of
the angle between the thrust axis of the selected \B\ candidate and
the thrust axis of the rest of the event
(all remaining charged and neutral candidates). The
distribution of \abscosttb\ is strongly peaked towards unity for
$\qqbar$ background whereas the distribution is uniform for signal events. 
We require \abscosttb\ $< 0.65$.
Additionally, we make requirements on a Fisher discriminant
$\cal{F}$~\cite{fisher} formed using a linear combination of
five variables. The first two variables are the momentum-weighted 
Legendre polynomial moments $L_0 = \sum_i p^*_i$ and 
$L_2 = \sum_i p^*_i \times |\cos(\theta^*_i)|^2$,
where $p^*_i$ is the momentum of particle $i$ (not from the \B\ candidate)
and $\theta^*_i$ is the angle
between its momentum and the thrust axis of the selected \B\ candidate in the
center-of-mass (CM) frame. We also use 
the absolute cosine of the angle between the direction of the \B\ and
the collision ($z$) axis in the CM frame, as well as the magnitude of the
cosine of the angle between the \B\ thrust axis
and the $z$ axis in the CM frame. The last
variable is the flavor of the recoiling \B\
as reported by a multivariate tagging algorithm~\cite{tagging}.
The selection requirements placed on \abscosttb\ and ${\cal{F}}$
are optimized using Monte Carlo simulated data and have a combined 
signal efficiency of 37\,\% while rejecting over 98\,\% of \qqbar\
background.

Other backgrounds arise from \BB\ events. The main
background for our charmless signal events 
is from charm decays, such as three- and four-body \B\ decays
involving an intermediate $D$ meson, and the charmonium decays
\JPsitoll\ and \Psitoll. We remove \B\ candidates when the invariant
mass of the combination of any two of its daughters tracks (of opposite charge)
is within the ranges $3.05 < m_{\pipi} < 3.22\gevcc$,  $3.68 < m_{\pipi} <
3.87\gevcc$ and  $1.70 < m_{\pipi} < 1.93\gevcc$, which reject the decays
\JPsitoll, \Psitoll\  and \DzbtoKpi\ (or \pipi), respectively.
These ranges are asymmetric about the nominal masses~\cite{pdg2004} in order to
remove decays in which a lepton ($\ell$) or kaon has been misidentified as a pion.

We study the remaining backgrounds from charmless \B\ decays and from charm
decays that escape the vetoes using a large sample of Monte
Carlo (MC) simulated \BB\ decays equivalent to approximately five
times the integrated luminosity for the data.  Any events that pass
the selection criteria are further studied using exclusive MC samples
to estimate reconstruction efficiency and yields.  
We find that the only significant background arises
from \BtoKPP\ decays, in which the kaon
has been misidentified as a pion.

We also consider the decay \BtoKspi, \Kstopipi\ to be a background, since 
the \KS\ candidates decay weakly and do
not interfere with other $\pi^+ \pi^-$ resonances in \BtoPPP.
We suppress this background
by fitting two oppositely charged pions from each \B\ candidate
to a common vertex when the invariant mass of the pair is below 0.6~\gevcc.
This vertex corresponds to the \KS\ decay point for true \Kstopipi\ candidates.
We remove \B\ decays that have fitted \KS\ candidates with masses between
476 and 519~\mevcc ($\sigma = 3.6~\mevcc$).

A further background in this analysis comes from signal events that
have been misreconstructed by switching one or more particles from the
decay of the signal \B\ meson with particles from the other \B\ meson
in the event. The amount of this background is estimated from MC
studies and is found to be very small; it accounts for 0.6\,\% 
of the final data sample in the signal region (defined in
Section~\ref{sec:finalsel}) and is therefore neglected in this analysis.

\section{Final Data Selection}
\label{sec:finalsel}

Two kinematic variables are used to select the final data sample. The
first variable is $\DeltaE = E_B^* - \sqrt{s}/2$, the difference
between the CM energy of the \B-meson candidate and
$\sqrt{s}/2$, where $\sqrt{s}$ is the total CM energy. The second is
the energy-substituted mass \mes\ $= \sqrt{(s/2 + {\bf p}_i \cdot {\bf
p}_B )^2/ E_i^2 - {\bf p}^2_B}$ where ${\bf p}_B$ is the $B$ momentum
and ($E_i$,{\bf p$_i$}) is the four-momentum of the initial state in
the laboratory frame.
For signal $B$ decays, the $\DeltaE$ distribution peaks near zero
with a resolution of 19~\mev, while the $\mes$ distribution peaks near the $B$
mass with a resolution of 2.7~\mevcc.
The mean of the \DeltaE\ distribution is shifted by $-5$~\mev\ from zero
in data as measured from the calibration channel \BtoDzbpi, \DzbtoKpi\, assuming
the kaon hypothesis for the $K^+$ candidate. The same shift is also observed
for \BtoDzbpi, \Dzbtopipi. The typical
\DeltaE\ separation between modes that differ by substituting a kaon
for a pion in the final state is 45~\mev, assuming the pion mass
hypothesis. Events in the \DeltaE\ strip $-65 < \DeltaE < 55$~\mev
are accepted. We also require events to lie in the range 
$5.20 < \mes < 5.29$~\gevcc.
This range is used for an extended maximum likelihood fit to the \mes\
distribution in order to determine the fraction of signal and background
events in our data sample. The region is further subdivided into two areas:
we use the sideband region ($5.20 < \mes < 5.26$~\gevcc) to study
the background Dalitz-plot distribution and the signal
region ($5.271 < \mes < 5.287$~\gevcc) to perform the Dalitz-plot
analysis. We accept one \B-meson candidate per event in the \DeltaE\ strip. Fewer
than 3\,\% of events have multiple candidates and in those events one
candidate is randomly accepted to avoid bias.
\begin{figure}[!htb]
\begin{center}
\includegraphics[angle=0,width=\columnwidth]{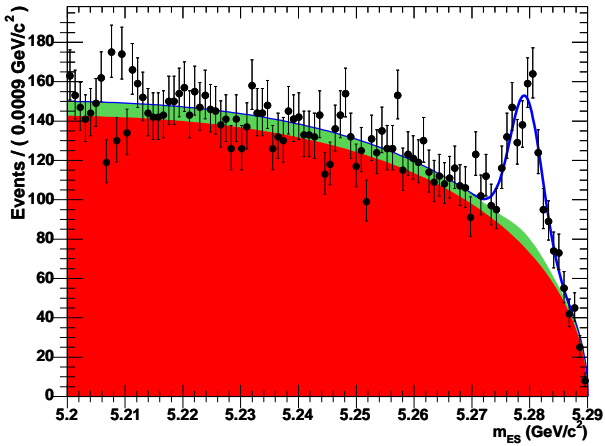}
\caption[\mes\ data and PDFs]
{The \mes\ distribution along the \DeltaE\ strip ($-65 < \DeltaE < 55$~\mev): 
the data are the black points with statistical error
bars, the lower solid red (dark) area is the \qqbar\ component, the middle solid green
(light) area is the \BB\ background contribution, while the upper
blue line shows the total fit result.}
\label{fig:mesfit}
\end{center}
\end{figure}

The \mes\ signal component is modeled by a Gaussian function, while the
\qqbar\ background is modeled using the ARGUS
function~\cite{argus} with the endpoint fixed to the beam energy while
the shape parameter is allowed to float. The \BB\ background \mes\ shape
is modeled with an ARGUS function plus a Gaussian to account for
the dominant peaking \BB\ background of $86 \pm 9$ \BtoKPP\ events, as well
as $7 \pm 1$ \BtoKspi\ events that have \KS\ candidates with
invariant masses outside the 6$~\sigma$ range.
All parameters of the \BB\ component,
including the amount of peaking and nonpeaking \BB\ background, are
obtained and fixed from the MC simulation. The fraction of \qqbar\ events
is allowed to float. Figure~\ref{fig:mesfit}
shows the \mes\ projection of the fit to the data for \BtoPPP.
The $\chi^2$ per degree of freedom for this projection is 93/95 and the
total number of events in the signal region is 1942 (965 and 977 for the
$B^-$ and $B^+$ samples, respectively).
In the signal region, the fraction of \qqbar\ background 
$f_{\qqbar}$ is found to be $(71.2 \pm 1.8)$\,\%,
while the fraction of \BB\ backgrounds $f_{\BB}$ is $(4.7 \pm 0.5)$\,\%. The 
fraction of signal events in the signal region is then
$f_{\rm{sig}} = 1 - f_{\qqbar} - f_{\BB} = (24.1 \pm 1.8)$\,\%.

\section{Dalitz Amplitude Analysis}
\label{sec:amplitude}

The charmless \B-meson decay to the final state \PPP\ has
a number of intermediate states in the Dalitz plot~\cite{Dalitz}
that contribute to the total rate, which can be represented in the form:

\begin{equation}
\frac{d\Gamma}{ds_{13} ds_{23}} = |{\cal{M}}|^2
\propto \left| \sum_k c_k e^{i\theta_k} {\cal{D}}_k(s_{13}, s_{23}) \right|^2
\end{equation}

\noindent where $s_{13} = m_{\pip \pim}^2$ and $s_{23} = m_{\pip \pim}^2$ are the 
invariant masses squared of the oppositely-charged pion pairs in the
final state. The invariant mass of each $B$ candidate is constrained
to the world-average value~\cite{pdg2004} before $s_{13}$
and $s_{23}$ are calculated. The amplitude for a given decay
mode $k$ is proportional to $c_k e^{i\theta_k} {\cal{D}}_k(s_{13}, s_{23})$ with magnitude
$c_k$ and phase $\theta_k$ ($-\pi \leq \theta_k \leq \pi$). 
The distributions ${\cal{D}}_k$ describe the dynamics of
the decay and are a product of the invariant mass and angular distributions.
For example, if we have a resonance formed from the first and third pion from \BtoPPP,
then
\begin{equation}
{\cal{D}}_k(s_{13}, s_{23}) = R_k(s_{13}) \times T_k(s_{13}, s_{23}),
\end{equation}
\noindent where $R_k(s_{13})$ is the resonance mass distribution and
$T_k(s_{13}, s_{23})$ is the angular-dependent amplitude. The ${\cal{D}}_k$ are
normalized such that
\begin{equation}
\int |{\cal{D}}_k(s_{13}, s_{23})|^2 ds_{13}ds_{23} = 1.
\end{equation}
The distribution $R_k(s_{13})$ is taken to be a
relativistic Breit--Wigner lineshape with 
Blatt--Weisskopf barrier factors~\cite{blatt}
for all resonances in this analysis except for the \fz, which is
modeled with a Flatt\'e lineshape~\cite{flatte} to account for
its coupled-channel behavior because it couples also to the $\Kp\Km$ channel 
right at threshold.
The nonresonant component is assumed to be uniform in phase space.  
The Breit--Wigner function has the form
\begin{equation}
R_k(s_{13}) = \frac{1}{m^2_0 - s_{13} - i m_0 \Gamma(s_{13})},
\label{eqn:BreitWigner}
\end{equation}
where $m_0$ is the nominal mass of the resonance and
$\Gamma(s_{13})$ is the mass-dependent width. In the general case, the latter
can be modeled as
\begin{equation}
\Gamma(s_{13}) = \Gamma_0 \left( \frac{q}{q_0}\right)^{2J+1} 
\frac{m_0}{\sqrt{s_{13}}} \frac{X^2_J(q)}{X^2_J(q_0)}.
\label{eqn:resWidth}
\end{equation}
The symbol $\Gamma_0$ denotes the nominal width of the resonance.
The values of $m_0$ and $\Gamma_0$ are obtained from standard tables~\cite{pdg2004}.
The value $q$ is the momentum of either daughter in the rest frame of the resonance,
and is given by
\begin{equation}
q = \sqrt{[s_{13} - (m^2_1 + m^2_3)][s_{13} - (m^2_1 - m^2_3)]/4s_{13}},
\label{eqn:qmomentum}
\end{equation}
where $m_1$ and $m_3$ are the masses of the two daughter particles, respectively.
The symbol $q_0$ denotes the value of $q$ when $s_{13} = m^2_0$. The Blatt--Weisskopf barrier
penetration factor $X_J(q)$ depends on the momentum $q$ as well as on the spin of the 
resonance $J$~\cite{blatt}:
\begin{eqnarray}
X_0(z) & = & 1 , \\
X_1(z) & = & \sqrt{1/(1 + z^2)} , \\
X_2(z) & = & \sqrt{1/(z^4 + 3z^2 + 9)} ,
\label{eqn:BlattEqn}
\end{eqnarray}
where $z = rq$ and $r$ is the radius of the barrier, which we take to be
4~GeV$^{-1}$ (equivalent to the approximate size of 0.8~fm).

In the case of the Flatt\'e lineshape~\cite{flatte}, 
which is used to describe the dynamics of the \fz\
resonance, the mass-dependent width is given by the sum 
of the widths in the $\pi\pi$ and $KK$ systems:
\begin{equation}
\Gamma(s_{13}) = \Gamma_{\pi}(s_{13}) + \Gamma_{K}(s_{13}) ,
\label{eqn:FlatteW1}
\end{equation}
where
\begin{eqnarray}
\Gamma_{\pi}(s_{13}) & = & g_{\pi} \sqrt{s_{13} - 4m^2_{\pi}} , \nonumber \\
\Gamma_{K}(s_{13}) & = & g_K \sqrt{s_{13} - 4m^2_{K}}
\label{eqn:FlatteW2}
\end{eqnarray}
and $g_{\pi}$ and $g_K$ are effective coupling constants, squared, for $\fz \ra \pi\pi$
and $\fz \ra KK$, respectively. We use the values $g_{\pi} = 0.138$
and $g_K = 4.45 g_{\pi}$ obtained by the BES collaboration~\cite{BESsigma}.

We use the Zemach tensor formalism~\cite{Zemach} for the angular distributions
$T^{(J)}_k$ of a spin 0 particle ($B^{\pm}$) decaying into a spin $J$ resonance
and a spin 0 bachelor particle ($\pipm$). For $J=0,1,2$, we have~\cite{Zemachpdg}:
\begin{eqnarray}
T^{(0)}_k & = & 1 , \nonumber \\
T^{(1)}_k & = & -2 \vec{p}\cdot\vec{q} , \nonumber \\
T^{(2)}_k & = & \frac{4}{3}\left[3(\vec{p}\cdot\vec{q})^2 - (|\vec{p}||\vec{q}|)^2\right],
\end{eqnarray}
where $\vec{p}$ is the momentum of the bachelor particle and $\vec{q}$
is the momentum of the like-sign resonance daughter, both measured
in the rest frame of the resonance.

To fit the data in the signal region, we define an unbinned likelihood
function for each event to have the form
\begin{eqnarray}
\label{LikeEqn}
& & {\cal{L}}(s_{13}, s_{23}) = f_{\rm{sig}} \nonumber \\
& & \frac{|\sum_{k=1}^{n} 
c_k e^{i\theta_k} {\cal{D}}_k(s_{13}, s_{23})|^2\epsilon(s_{13}, s_{23})}
{\int~{|\sum_{k=1}^{n} c_k e^{i\theta_k} {\cal{D}}_k(s_{13}, s_{23})|^2\epsilon(s_{13}, s_{23})}~ds_{13}ds_{23}} \nonumber \\
& & +~f_{q\bar{q}}~\frac{Q(s_{13}, s_{23})}
{\int~Q(s_{13}, s_{23})~ds_{13}ds_{23}} \nonumber \\
& & +~f_{B\bar{B}}~\frac{B(s_{13}, s_{23})}
{\int~B(s_{13}, s_{23})~ds_{13}ds_{23}}
\end{eqnarray}
where $n$ is the total number of resonant and nonresonant components in the signal model;
$\epsilon(s_{13},s_{23})$ is the signal reconstruction efficiency defined for all points in the
Dalitz plot; $Q(s_{13},s_{23})$ is the distribution of \qqbar\ background; $B(s_{13},s_{23})$
is the distribution of \BB\ background; and $f_{\rm{sig}}$, $f_{\qqbar}$ and $f_{\BB}$ are the 
fractions of signal, \qqbar\ and \BB\ backgrounds, respectively.
Since we have two identical pions in the final state,
the dynamical amplitudes, signal efficiency
and background distributions are symmetrized between $s_{13}$ and $s_{23}$.
The fit is performed allowing the amplitude magnitudes ($c_i$) and the
phases ($\theta_i$) to vary.

The first term on the right-hand-side in Eq.~(\ref{LikeEqn}) corresponds
to the signal probability density function (PDF) multiplied by the
signal fraction $f_{\rm{sig}}$. 
This analysis will only be sensitive to relative phases
and magnitudes, since we can always apply a common magnitude scaling
factor and phase transformation to all terms in the numerator and denominator
of the signal PDF. Therefore, we have fixed the magnitude and phase of the 
most dominant component, \rhoI, to be 1 and 0, respectively.

As the choice of normalization, phase convention and amplitude
formalism may not always be the same for different experiments, fit
fractions are also presented to allow a
more meaningful comparison of results. The fit fraction for
resonance $k$, $F_k$, is defined as
the integral of a single decay amplitude squared divided by the
coherent matrix element squared for the complete Dalitz plot as shown
in Eq.~(\ref{eqn:fitfraction}).
\begin{equation}
F_k = 
\frac{\int|c_k e^{i\theta_k}{\cal{D}}_k(s_{13}, s_{23})|^2ds_{13}ds_{23}}
{\int |\sum_j c_j e^{i\theta_j} {\cal{D}}_j(s_{13}, s_{23})|^2 ds_{13}ds_{23}},
\label{eqn:fitfraction}
\end{equation}
where the integrals are performed over the full kinematic range.
Note that the sum of these fit fractions is not necessarily unity due
to the potential presence of net constructive or destructive
interference.

\section{Dalitz Plot Backgrounds and Efficiency}
\label{sec:bgandeff}

The dominant source of background for this analysis comes from
\qqbar\ events. We use a combination of on-resonance sideband
data and off-resonance data to get the background
distribution for the Dalitz plot. Note that for the on-resonance sideband data,
we subtract any contributions from $\BB$ background (from MC), since this
is handled separately. Since the background peaks at the edges of the
Dalitz plot, we use a coordinate transformation to a square Dalitz plot
in order to improve the modeling of the background distribution.
Considering the decay $B^+ \ra \pi^+ \pi^+ \pi^-$, 
the new coordinates are $m'$ and $\theta'$, which are defined as

\begin{eqnarray}
\label{eqn:squareDP}
m' & = & \frac{1}{\pi} \cos ^{-1}\left(2 \frac{m_{++} - m_{++}\rm{[min]}}
{m_{++} {\rm{[max]}} - m_{++} \rm{[min]}} - 1\right) , \nonumber \\
\theta' & = & \frac{1}{\pi}\theta_{++},
\end{eqnarray}
where $m_{++}$ is the invariant mass of the like-sign
pions, $m_{++}{\rm{[max]}} = m_B - m_{\pi}$ and $m_{++}{\rm{[min]}} = 2m_{\pi}$
are the boundaries of $m_{++}$, while $\theta_{++}$ is the helicity
angle between the momentum of one of the like-sign ($\pi^+$) pions and the $\pi^-$ momentum
in the $\pi^+ \pi^+$ rest-frame. 
Note that the new variables range from 0 to 1. The Jacobian transformation $J$ 
between the normal Dalitz plot variables to the new coordinates is defined as
\begin{equation}
\label{eqn:Jacobian}
ds_{13} ds_{23} = |J| dm' d\theta'.
\end{equation}
The determinant $|J|$ of the Jacobian is given by
\begin{equation}
\label{eqn:detJ}
|J| = 4 |p_1^*| |p_2^*| m_{++} \frac{\partial m_{++}}{\partial m'} 
\frac{\partial \cos\theta_{++}}{\partial \theta'},
\end{equation}
where $|p^*_1|$ is the momentum of one of the $\pi^+$ candidates and $|p^*_2|$ is the
momentum of the $\pi^-$ track, both measured in the rest frame of the $\pi^+ \pi^+$ system. 
The partial derivatives in Eq.~(\ref{eqn:detJ}) are given by
\begin{eqnarray}
\label{eqn:derivatives}
\frac{\partial m_{++}}{\partial m'} & = & -\frac{\pi}{2} \sin (\pi m')
(m_{++} {\rm{[max]}} - m_{++} {\rm{[min]}} ), \nonumber \\
\frac{\partial \cos \theta_{++}}{\partial \theta'} & = & -\pi \sin (\pi \theta').
\end{eqnarray}

\begin{figure*}[!htb]
\begin{center}
\includegraphics[width=\textwidth]{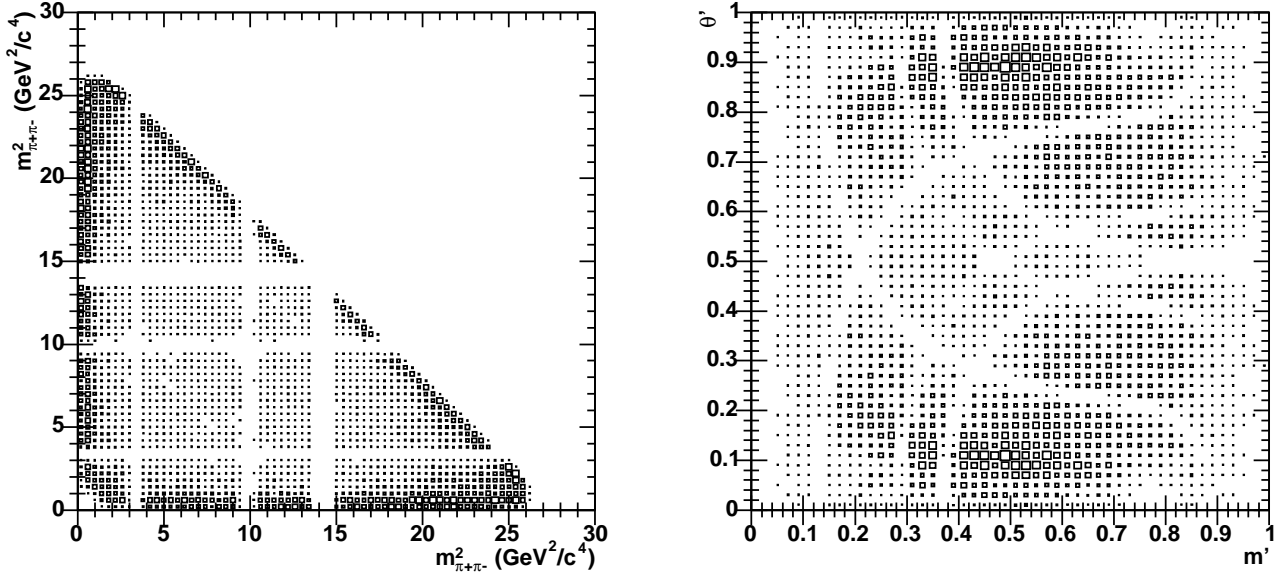}
\caption[Dalitz plot of the \qqbar\ background obtained
from on-resonance sideband and off-resonance data.]
{Dalitz plot of the \qqbar\ background obtained
from on-resonance sideband and off-resonance data. The left plot
shows the distribution in normal Dalitz plot coordinates, while the right
plot shows the equivalent distribution in the new square Dalitz plot
coordinates, defined in Eq.~(\ref{eqn:squareDP}). The empty regions
correspond to events removed by the charm vetoes. The area of each small square is 
proportional to the number of events in that bin.}
\label{fig:BgDPPlot}
\end{center}
\end{figure*}

\begin{figure*}[!hbt]
\begin{center}
\includegraphics[width=\textwidth]{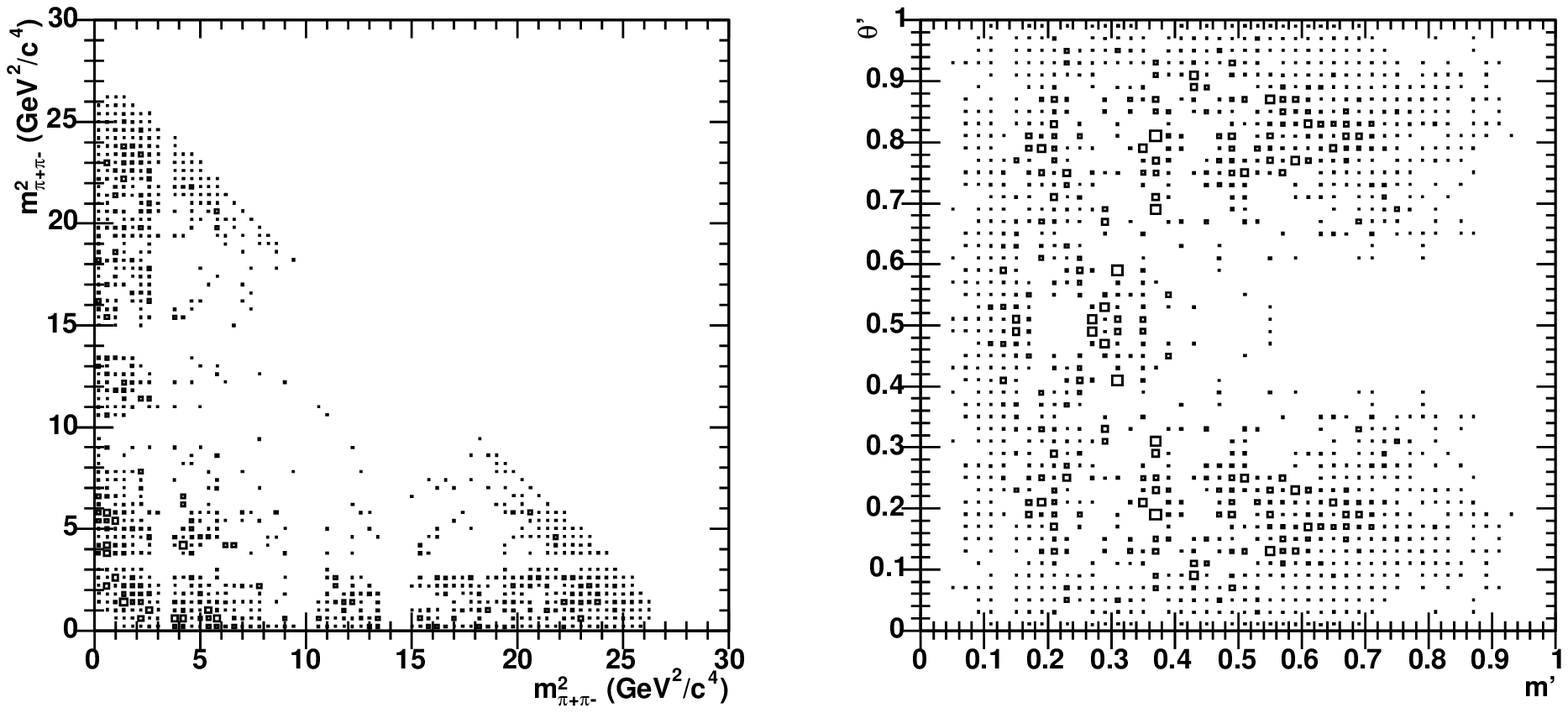}
\caption[Dalitz plot of the $\BB$ background obtained
from Monte Carlo simulated events.]
{Dalitz plot of the $\BB$ background obtained
from Monte Carlo simulated events. The left plot
shows the distribution in normal Dalitz plot coordinates, while the right
plot shows the equivalent distribution in the new square Dalitz plot
coordinates, defined in Eq.~(\ref{eqn:squareDP}). The empty regions
correspond to events removed by the charm vetoes. The area of each small square is 
proportional to the number of events in that bin.}
\label{fig:BBbarDPPlot}
\end{center}
\end{figure*}

\noindent
We get similar expressions for $B^- \ra \pi^- \pi^- \pi^+$. 
Figure~\ref{fig:BgDPPlot} shows the \qqbar\ background distribution,
obtained by combining on-resonance sideband and off-resonance data.
Figure~\ref{fig:BBbarDPPlot} shows the \BB\ background distributions, which 
originate from \BtoKPP\ and \BtoKspi\ decays.
Note that
the peaks along the edges of the normal Dalitz plot distribution are more spread
out in the square Dalitz plot format. We use the latter to represent the \qqbar\
and \BB\ backgrounds in the amplitude fit, applying linear interpolation between
bins.

The signal efficiency $\epsilon(s_{13},s_{23})$ used in 
Eq.~(\ref{LikeEqn}) is modeled using a two-dimensional
histogram with bins of size $0.4~(\gevcc)^2 \times 0.4~(\gevcc)^2$ 
and is obtained using 1.1 million \BtoPPP\ nonresonant MC events. All selection criteria
are applied except for those corresponding to the invariant-mass veto regions
mentioned in Sec.~\ref{sec:background}.
The efficiency at a given bin is defined as the ratio of the
number of events reconstructed to the number of events generated in that bin.
Corrections for differences between MC and data in the particle
identification and tracking efficiencies are applied. The efficiency
shows little variation across the majority of the Dalitz plot, in which
the average efficiency is measured to be $(13.00 \pm 0.04)\,\%$, however there are
decreases towards the corners where one of the particles has a low momentum.
The effect of experimental resolution on the signal model is neglected
since the resonances under consideration are sufficiently broad.
No difference in efficiency is seen between $B^-$ and $B^+$ decays at the 2\% level.

\section{Physics Results}
\label{sec:Physics}

We fit the $B^-$ and $B^+$ samples independently to extract
the magnitudes and phases of the resonant and nonresonant
contributions to the charmless \BtoPPP\ Dalitz plot, using Eq.~(\ref{LikeEqn}).
The nominal fit model contains the resonances \rhoI, \rhoII, \fz,
\fzII\ and a uniform nonresonant contribution. This is chosen using
information from established resonance states~\cite{pdg2004}
and the $\chi^2$ variation observed when omitting one of the five
components. The $\chi^2$ value is calculated using the formula
\begin{equation}
\chi^2 = \sum_{i=1}^{n_b} \frac{[y_i - f(x_i)]^2}{\sigma_i^2},
\label{eqn:chisq}
\end{equation}
where $y_i$ is the number of events in bin $i$ of the 
invariant mass or Dalitz-plot distribution, $f(x_i)$ is the expected
number of events in that bin as predicted by the fit result and
$\sigma_i$ is the error on $y_i$ ($\sqrt{y_i}$). The number of degrees of freedom (nDof)
is calculated as $n_b - k - 1$, where $n_b$ is the total number of bins used
and $k$ is the number of free parameters in the fit (4 magnitudes and 4 phases).
A minimum of 10 entries in each bin is required; if this requirement is not met
then neighboring bins are combined.
Typically, $n_b$ is equal to 35 and 75 for the invariant mass and Dalitz-plot distributions, 
respectively. Since we observe no charge asymmetry in the \qqbar\ and \BB\ backgrounds,
we use the charge-averaged background distributions
shown in Figs.~\ref{fig:BgDPPlot} and~\ref{fig:BBbarDPPlot} for 
the $B^-$ and $B^+$ fits.
The results of the nominal fit to $B^-$ and $B^+$ on-resonance data 
in the signal region are shown separately in Table~\ref{tab:pipipi_nominal}.
\begin{table*}[!htb]
\caption[Results of the nominal fit for the $\BtoPPP$ mode.]
{Results of the nominal fits to \Btopppneg\ and \Btoppppos\ data.
The first errors are statistical, while the second and third uncertainties are systematic
and model-dependent, respectively, all of which are detailed in Sec.~\ref{sec:Systematics}.
All phases are in radians.}
\label{tab:pipipi_nominal}
\begin{center}
\begin{tabular}{lcccc}
\hline
Component  & & $B^-$ Fit Result & & $B^+$ Fit Result\\
\hline
\hline
$\rhoI$  Fraction (\%)     & & $50.6 \pm 7.3 \pm 2.2^{+0.4}_{-2.7}$ & & $57.8 \pm 6.8 \pm 3.5^{+1.0}_{-7.9}$ \\
$\rhoI$  Magnitude         & & 1.0~(fixed) & & 1.0~(fixed) \\
$\rhoI$  Phase             & & 0.0~(fixed) & & 0.0~(fixed) \\
\hline
$\rhoII$  Fraction (\%)    & & $6.8 \pm 4.5 \pm 1.8 \pm 0.6$ & & $4.9 \pm 5.5 \pm 1.4^{+0.8}_{-2.3}$ \\
$\rhoII$ Magnitude         & & $0.37 \pm 0.11 \pm 0.05 \pm 0.02$ & & $0.29 \pm 0.17 \pm 0.06 \pm 0.08$ \\
$\rhoII$ Phase             & & $+1.99 \pm 0.57 \pm 0.10 \pm 0.08$ & & $+0.31 \pm 0.70 \pm 0.15 \pm 0.34$ \\
\hline
$\fz$  Fraction (\%)       & & $3.8 \pm 4.9 \pm 0.9 \pm 2.1$ & & $11.1 \pm 5.1 \pm 1.1^{+4.4}_{-3.4}$ \\
$\fz$ Magnitude            & & $0.27 \pm 0.10 \pm 0.05 \pm 0.07$ & & $0.44 \pm 0.12 \pm 0.03 \pm 0.08$ \\
$\fz$ Phase                & & $-1.59 \pm 0.47 \pm 0.08^{+0.15}_{-0.01}$ & & $-0.79 \pm 0.62 \pm 0.17^{+0.15}_{-0.02}$ \\
\hline
$\fzII$  Fraction (\%)     & & $14.2 \pm 4.6 \pm 1.3 \pm 0.5$ & & $14.1 \pm 4.8 \pm 1.4^{+0.7}_{-3.1}$ \\
$\fzII$ Magnitude          & & $0.53 \pm 0.10 \pm 0.02 \pm 0.03$ & & $0.49 \pm 0.11 \pm 0.02 \pm 0.05$ \\
$\fzII$ Phase              & & $+1.39 \pm 0.41 \pm 0.09 \pm 0.09$ & & $+1.85 \pm 0.47 \pm 0.12^{+0.39}_{-0.07}$ \\
\hline
Nonresonant  Fraction (\%) & & $15.0 \pm 8.6 \pm 1.9^{+4.3}_{-1.3}$ & & $12.6 \pm 7.1 \pm 2.6^{+1.1}_{-4.3}$ \\
Nonresonant Magnitude      & & $0.54 \pm 0.13 \pm 0.03 \pm 0.09$ & & $0.47 \pm 0.14 \pm 0.05 \pm 0.06$ \\
Nonresonant Phase          & & $-0.84 \pm 0.38 \pm 0.06 \pm 0.04$ & & $-2.80 \pm 0.46 \pm 0.07 \pm 0.07$ \\
\hline
\end{tabular}
\end{center}
\end{table*}
From Eq.~(\ref{eqn:fitfraction}), it can be seen that the
fit fraction statistical uncertainty will not
only depend on the uncertainties of the magnitude and phase
of the given resonance, but also on the statistical errors of all amplitudes.
Therefore, we use a MC pseudo-experiment technique to obtain the statistical
uncertainty on each fit fraction. Each pseudo-experiment is a sample of MC
generated events that contains the correct mixture of signal and background, 
which are distributed across the Dalitz plot 
according to the PDFs defined in Eq.~(\ref{LikeEqn}). We fit these MC samples
and plot the distributions of fit fractions $F_k$ obtained from a thousand such experiments.
The statistical uncertainty for each $F_k$ is then the value of the
width of the Gaussian function that is fitted to the $F_k$ distribution.

\begin{figure*}[!htb]
\begin{center}
\includegraphics[width=1.0\textwidth]{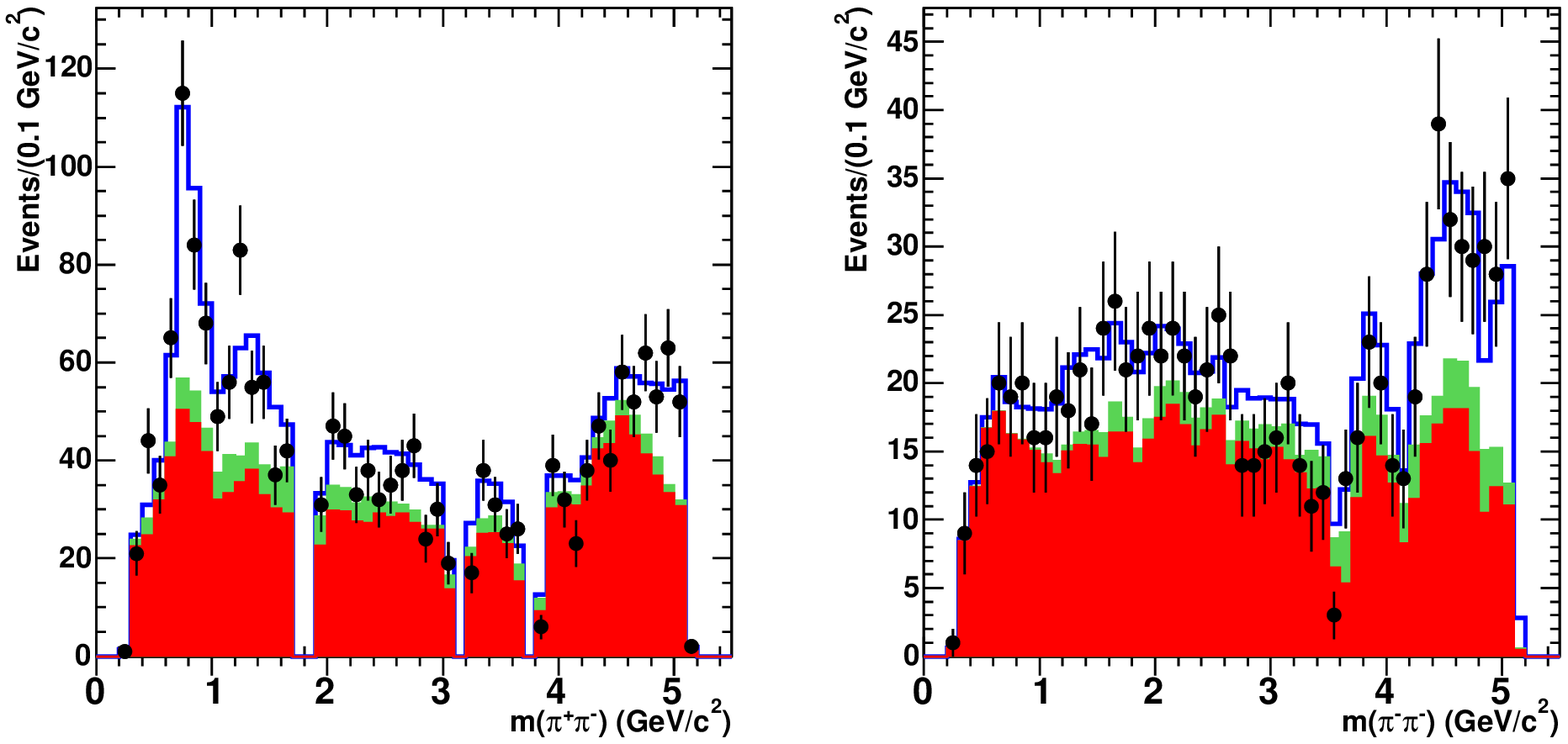}
\includegraphics[width=1.0\textwidth]{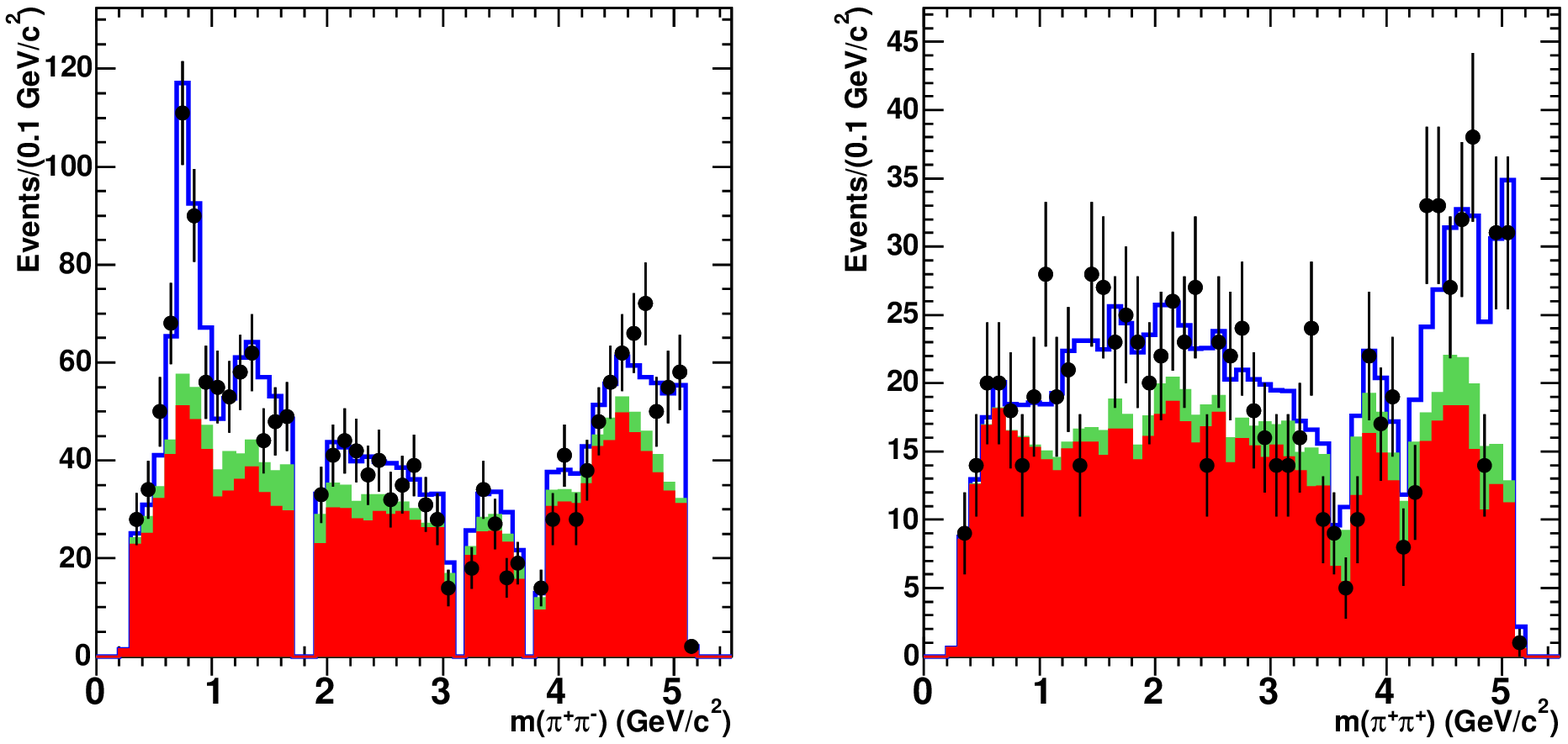}
\caption{Projection plots of the fit results for \Btopppneg\ and \Btoppppos\ 
onto the mass variables $m_{\pi^+ \pi^-}$ and $m_{\pi^- \pi^-}$ ($m_{\pi^+ \pi^+}$).
The upper (lower) plots are for the $B^-$ ($B^+$) data sample.
The data are the black points with statistical error bars, the lower solid red (dark) 
histogram is the \qqbar\ component, the middle solid green (light) histogram is the
\BB\ background contribution, while the upper blue histogram shows
the total fit result. The large dips in the spectra correspond to the charm vetoes.}
\label{fig:Fit3piData}
\end{center}
\end{figure*}

\begin{figure}[!htb]
\begin{center}
\includegraphics[width=\columnwidth]{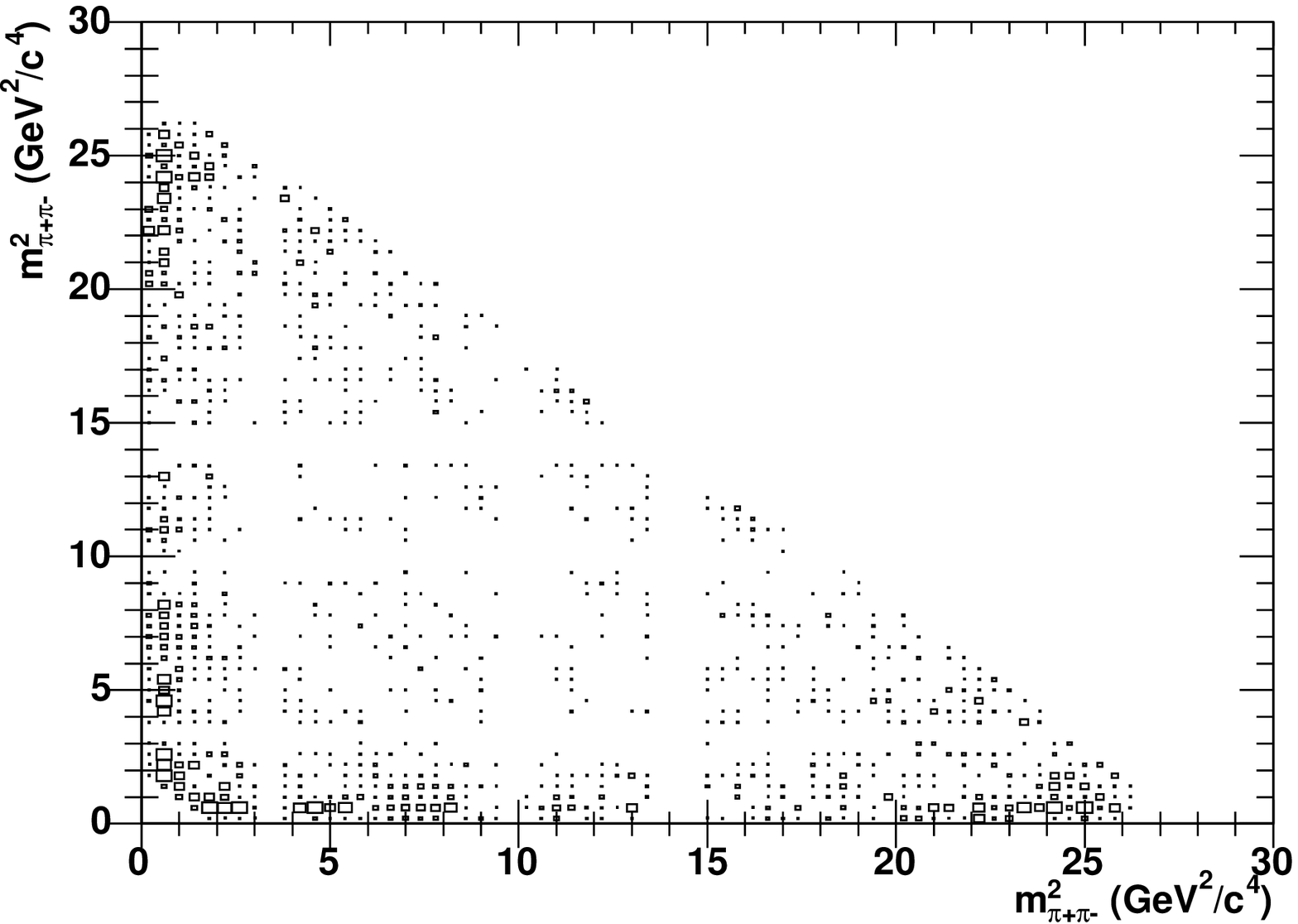}
\caption{Background subtracted Dalitz plot of the combined \BtoPPP\ data
sample in the signal region. The Dalitz plot is symmetrized about the $y=x$ axis.
The empty regions correspond to events removed by the charm vetoes.}
\label{fig:onResDP}
\end{center}
\end{figure}

Figure~\ref{fig:Fit3piData} shows the mass projection plots for the
nominal fits to $B^-$ and $B^+$ data, while Fig.~\ref{fig:onResDP}
shows the background-subtracted Dalitz plot of the combined \BtoPPP\ data in the signal region. 
The $\chi^2$/nDof values for the
opposite-sign and like-sign invariant mass projections for $B^-$ ($B^+$) are 
51/34 and 27/37 (35/35 and 47/35), respectively. The
$\chi^2$/nDof values for the two-dimensional Dalitz plots are 74/74
and 70/75 for $B^-$ and $B^+$, respectively.
The four resonant contributions plus the single uniform
phase-space nonresonant model are able to describe the data adequately
within the statistical uncertainties.
For a given resonance, the comparison of the fit fraction, not the magnitude, to its
uncertainty gives a measure of how significant its contribution is to the Dalitz plot.
Note that the fit fraction uncertainties shown in 
Table~\ref{tab:pipipi_nominal} are larger than the uncertainties
of the magnitudes. This is due to the dependence of the former
on all of the other amplitudes, via the denominator in Eq.~(\ref{eqn:fitfraction}).
It can be clearly seen that the dominant contribution to the 
charmless \BtoPPP\ Dalitz plot is from the \rhoI\
resonance. Approximately 10\,\% of the \rhoI\ fit fraction
lies in the tail of the mass distribution, defined as the 
region outside $m_0 \pm 3\Gamma_0$. In addition, the
fraction of \rhoI\ within one width of the \fzII\ resonance line-shape
is approximately 13\,\%, which is equivalent to half of the \fzII\ fit fraction.
Further fits are performed to the data by removing one two-body
component at a time from the nominal model. Removing the \rhoI, \fzII\
or nonresonant components give significantly poorer fit results. 
Omitting the \rhoII\ or \fz\ components, which are present at the $1.5~\sigma$ level,
gives a small change in the goodness-of-fit $\chi^2$ (Eq.~(\ref{eqn:chisq})).

We have also tested the introduction of the \chiczero\ and \fzIII\ 
resonances, as well as the low-mass $\pi^+ \pi^-$ pole, 
known as the $\sigma$.
Analysis of data from the E791 experiment for 
$D^+ \ra \pi^+ \pi^- \pi^+$~\cite{E791} and recent data from the BES collaboration
for $J/\psi \ra \omega \pi^+ \pi^-$~\cite{BESsigma} show evidence
of the $\sigma$. Also
a large concentration of events in the $I=0$ S-wave $\pi \pi$ channel
has been seen in the $m_{\pi \pi}$ region around $500 - 600$~MeV in
$pp$ collisions~\cite{Alde}. This pole is predicted from models based
on chiral perturbation theory~\cite{Colangelo}, in which the resonance
parameters are $M - i\Gamma/2 = \left[(470 \pm 30) - i(295 \pm 20)\right]$~MeV.
Consequently, the $\sigma$ resonance is predicted in \BtoPPP\ decays.
For this Dalitz-plot analysis the $\sigma$ resonance is modeled using the 
parameterization suggested by Bugg~\cite{bugg}. 
The contributions that these three resonances
make to the nominal fit results are not significant and so we place upper limits on
them.

To make comparisons with previous measurements and theoretical predictions
it is necessary to convert the fit fractions into branching fractions.
These are estimated by multiplying each fit fraction by the total branching fraction
for the $B^-$ and $B^+$ fits, which are then averaged. The total branching fractions
${\cal{B}}^-_{\rm{tot}}$ and ${\cal{B}}^+_{\rm{tot}}$ 
for $\Btopppneg$ and $\Btoppppos$, respectively, are defined as
\begin{equation}
{\cal{B}}_{\rm{tot}}^{\pm} = \frac{N^{\pm} f_{\rm{sig}}}{N_{\BB} \left<\epsilon^{\pm}\right>},
\label{BFEqn}
\end{equation}
where $N^{\pm}$ is the total number of events in the signal region, $f_{\rm{sig}}$
is the signal fraction defined earlier, $N_{\BB}$ is half the
total number of \BB\ pairs in the data sample~\cite{BBbarPairs} 
and $\left<\epsilon^{\pm}\right>$
is the average efficiency across the Dalitz plot weighted by the
fitted signal distribution, which is equal to $(12.4 \pm 0.1)$\,\% for $B^-$ and $B^+$.
The average total branching fraction is then just
equal to $\frac{1}{2}({\cal{B}}^+_{\rm{tot}} + {\cal{B}}^-_{\rm{tot}})$, while
the average branching fraction for each resonance $k$ is given by
\begin{equation}
{\cal{B}}_k = \frac{1}{2} (F_k^- {\cal{B}}^-_{\rm{tot}} + F_k^+ {\cal{B}}^+_{\rm{tot}}),
\label{ResBFEqn}
\end{equation}
where $F^-_k$ ($F^+_k$) is the fit fraction for resonance $k$ for $B^-$ ($B^+$).

For components that do not have
statistically significant fit fractions 90\,\% confidence-level upper limits are evaluated. 
Upper limits are also found for the \chiczero, \fzIII\ and $\sigma$ components.
These limits are calculated by generating many pseudo-MC experiments from the
results of fits to the data, with all systematic sources 
(see Sec.~\ref{sec:Systematics}) varied within their $1~\sigma$ Gaussian uncertainties.
We fit these MC samples and plot the fit fraction distributions.
The 90\,\% confidence-level upper limit for each fit fraction is then that which removes 90\,\% of
the pseudo-MC experiments. A branching fraction upper limit is then the product
of the upper limit on a fit fraction with the total branching fraction
${\cal{B}}^{\pm}_{\rm{tot}}$. 
Corrections applied to the signal efficiency due to
differences between data and MC are described in Sec.~\ref{sec:Systematics}. 
We include the variation of $\left<\epsilon^{\pm}\right>$ due to these corrections
by using another large set of pseudo-MC experiments, which 
is generated and fitted to the Dalitz-plot
model. The content of each bin in the efficiency histogram is increased (decreased) 
by the same random
fluctuation given by the uncertainty of the efficiency correction (5.1\,\%).
The 90\,\% confidence-level upper limit on the value of the reciprocal of the efficiency
(1/0.117) is taken as the value of
$1/\left<\epsilon^{\pm}\right>$ for the total branching fraction calculation
given in Eq.~(\ref{BFEqn}) that is then used to find the upper limits for the
resonance branching fractions. If the upper limits differ between $B^-$ and $B^+$, we
choose the larger value to be conservative.

In addition to fit fractions and phases, the charge ($\CP$) asymmetries
for the signal model components are also measured.
The charge asymmetry for the total branching fraction is defined as
\begin{equation}
\label{eqn:CPAsymm}
{\cal{A}} = \frac{N^-_{\rm{sig}} - N^+_{\rm{sig}}}{N^-_{\rm{sig}} + N^+_{\rm{sig}}},
\end{equation}
where $N^-_{\rm{sig}}$ ($N^+_{\rm{sig}}$) is the number of signal events for
the $B^-$ ($B^+$) sample. The charge asymmetries for the fit fractions are defined as
\begin{equation}
\label{eqn:FitFracAsymm}
{\cal{A}}_k = \frac{F^-_k N^-_{\rm{sig}} - F^+_k N^+_{\rm{sig}}}
{F^-_k N^-_{\rm{sig}} + F^+_k N^+_{\rm{sig}}}.
\end{equation}
The measured branching fractions and charge asymmetries are summarized in
Table~\ref{tab:BFCPResults}. The total branching fraction of the charmless
\BtoPPP\ decay, $(16.2 \pm 1.2 \pm 0.9) \times 10^{-6}$, 
is consistent with the current world-average value of 
$(11 \pm 4) \times 10^{-6}$~\cite{pdg2004}. The measured branching fraction
for the decay $B^{\pm} \ra \rhoI \pi^{\pm}, \rhoI \ra \pi^+ \pi^-$,
$(8.8 \pm 1.0 \pm 0.6^{+0.1}_{-0.7}) \times 10^{-6}$,
agrees with the world-average value of $(8.6 \pm 2.0) \times 10^{-6}$~\cite{pdg2004}
and is consistent with the average theoretical predictions
of $11.9 \times 10^{-6}$ and $8.4 \times 10^{-6}$ that are 
based on QCD factorization~\cite{neubert} and pole-dominance 
models~\cite{kramer}, respectively.
The upper limits reported for the other resonance modes are an order
of magnitude lower than published limits~\cite{pdg2004}. The total charge asymmetry
has been measured to be consistent with zero to a higher degree of accuracy than
previous measurements~\cite{Babar2}.
A representative theoretical value of the charge asymmetry 
for $B^{\pm} \ra \rhoI \pi^{\pm}$ is +4.1\,\%~\cite{neubert}, ignoring uncertainties due to weak
annihilation processes, in agreement with our measurement.

\begin{table*}[!htb]
\caption{Summary of average branching fraction (${\cal{B}}$) and charge asymmetry 
(${\cal{A}}$) results. The first
uncertainty is statistical, the second is systematic, while the third
is model-dependent.}
\label{tab:BFCPResults}
\begin{center}
\begin{tabular}{lccccc}
\hline
Mode & & ${\cal{B}}(B^{\pm} \ra {\rm{Mode}}) (10^{-6})$ & & 90\% CL UL ${\cal{B}}$ $(10^{-6})$ & ${\cal{A}}$ (\%) \\
\hline
\BtoPPP\ Total & & $16.2 \pm 1.2 \pm 0.9$ & & --- & $-0.7 \pm 7.7 \pm 2.5$ \\
\hline
$\rhoI \pi^{\pm}, \rhoI \ra \pip \pim$ & & $8.8 \pm 1.0 \pm 0.6^{+0.1}_{-0.7}$ & & --- & $-7.4 \pm 12.0 \pm 3.4^{+0.6}_{-4.4}$ \\
$\rhoII \pi^{\pm}, \rhoII \ra \pip \pim$ & & $1.0 \pm 0.6 \pm 0.2 \pm 0.2$ & & $< 2.3$ & $+15.5 \pm 62.1 \pm 7.9^{+0.4}_{-1.0}$ \\
$\fz \pi^{\pm}, \fz \ra \pip \pim$ & & $1.2 \pm 0.6 \pm 0.1 \pm 0.4$ & & $< 3.0$ & $-49.5 \pm 53.7 \pm 4.9^{+3.7}_{-2.9}$ \\
$\fzII \pi^{\pm}, \fzII \ra \pip \pim$ & & $2.3 \pm 0.6 \pm 0.2 \pm 0.3$ & & $< 3.5$ & $-0.4 \pm 24.7 \pm 2.8^{+0.4}_{-1.6}$ \\
\BtoPPP\ Nonresonant & & $2.3 \pm 0.9 \pm 0.3 \pm 0.4$ & & $< 4.6$ & $+8.0 \pm 41.2 \pm 6.5 \pm 2.4$ \\
$\chiczero \pi^{\pm}, \chiczero \ra \pip \pim$ & & --- & & $< 0.3$ & --- \\
$\fzIII \pi^{\pm}, \fzIII \ra \pip \pim$ & & --- & & $< 3.0$ & --- \\
$\sigma \pi^{\pm}, \sigma \ra \pip \pim$ & & --- & & $< 4.1$ & --- \\
\hline
\end{tabular}
\end{center}
\end{table*}

\section{Systematic Studies}
\label{sec:Systematics}

The systematic uncertainties that affect the measured fit fractions,
amplitude magnitudes and phases are evaluated separately for $B^-$ and $B^+$.
The first source of systematic uncertainty
is the modeling of the signal efficiency.
The charged-particle tracking and particle-identification
fractional uncertainties are 2.4\,\% and 4.2\,\%, respectively. 
The first is estimated by finding the difference between data and MC
of the track-finding efficiency of the DCH from multihadron events.
A precise determination of the DCH efficiency can be made
by observing the fraction of tracks in the
SVT that are also found in the DCH. The probabilities
of identifying kaons and pions is measured using the decay mode
$D^{*+} \ra D^0 \pi^+, D^0 \ra K^- \pi^+$, which
provides a very pure sample of pions and kaons. The difference
observed between data and MC for the kaon and pion efficiencies gives
the combined systematic uncertainty of 4.2\,\% for our signal mode.
There are also global systematic errors in the efficiencies due to the criteria
applied to the event-shape variables (1.0\,\%) and to $\Delta E$ and
$m_{ES}$ (1.0\,\%). The total fractional systematic uncertainty
for the efficiency from these sources is 5.1\,\%. Corrections due to differences
between data and MC have also been included for the selection
requirements on \costtb, ${\cal{F}}$, \DeltaE and \mes.
These are found by comparing the difference in the selection efficiency
between data and MC for the control sample \BtoDzbpi.

The variation of the efficiency across the Dalitz plot
is also evaluated by performing a series of fits to the data
where the efficiency histogram has each bin fluctuate
in accordance with its binomial error. This introduces
an absolute uncertainty of 0.01 for the magnitudes, 0.02 to 0.05 for the phases, and
a fractional uncertainty between 1\,\% and 4\,\% for the fit fractions.
For the average efficiency, and hence for the total branching fraction,
this is a very small effect, evaluated at 0.1\,\%.

The next source of systematic uncertainty comes from the modeling
of the backgrounds.
The systematic uncertainty introduced by the \BB\ background and
\qqbar\ background has two components, each of which can potentially
affect the fitted magnitudes and phases differently. The first
component arises from the uncertainty in the overall normalization of
these backgrounds, while the second component arises from the
uncertainty on the shapes of the background distributions in the
Dalitz plot. The uncertainties on the magnitudes, phases and fit fractions due to
the normalization uncertainty are estimated by varying the measured
background fractions in the signal region by their statistical errors.
The maximum uncertainty for the magnitude (phase) is 0.03 (0.02)
due to the \qqbar\ background normalization uncertainty and
0.01 (0.01) due to the \BB\ background normalization uncertainty. These
uncertainties are added in quadrature. The fit fractions have
relative uncertainties in the range
1\,\% to 9\,\%. The uncertainties on the fit fractions and
phases due to the Dalitz-plot background distribution uncertainty is
estimated in the same way as the efficiency variation, namely varying
the contents of the histogram bins in accordance with their Poisson
errors. To be conservative, each magnitude (phase) has 
been given an
uncertainty of 0.02 (0.02) due to the \qqbar\ background distribution
uncertainty and 0.02 (0.01) due to the \BB\ background distribution
uncertainty, which are then added in quadrature.  
The fit fractions have relative uncertainties ranging from 1\,\% to 10\,\%.

To confirm the fitting procedure, 1000 MC pseudo-experiments
are created from the fitted magnitudes and phases and each sample is fitted
100 times with randomized starting parameters.
A fit bias of approximately 10\,\% is observed for some of the smaller components and is included
in the systematic uncertainties for the magnitudes, phases and fit fractions.

There is a range of different values for the coupling constants $g_{\pi}$
and $g_{K}$ for the Flatt\'e description of the \fz\ resonance~\cite{BESsigma,WA76,E791Flatte}.
A model-dependent systematic uncertainty is assigned for all magnitudes, phases
and fit fractions based on the differences between the results of the nominal fit
and those when the different coupling constants for the \fz\ are used.
There is also the question of whether the nonresonant
component has an amplitude that varies across the Dalitz plot. 
For the nominal fit,
uniform phase-space is used for this component in the absence of any
{\it{a priori}} model. An alternative parameterization gives 
the nonresonant dynamical amplitude to be of the form
\begin{equation}
{\cal{D}}_{NR}(s_{13}, s_{23}) = e^{-\alpha s_{13}} + e^{-\alpha s_{23}},
\label{eqn:BelleNR}
\end{equation}
where $\alpha$ is a constant~\cite{BelleNR}. This parameterization does not give
significant differences compared to the nominal fit results (Table~\ref{tab:pipipi_nominal}) for
$\alpha = 0.11 \pm 0.02$, which is the average
of the values shown in Ref.~\cite{BelleNR}. These differences are included in the model-dependent
systematic error, as well as when the $\chiczero$, $\fzIII$ or $\sigma$ resonances are added to the fit.

The dominant systematic uncertainty for the total branching fraction 
${\cal{B}}^{\pm}_{\rm{tot}}$ is due to the efficiency corrections (5.1\,\%). There is a 1\,\% 
fractional error on the weighted efficiency 
$\left<\epsilon\right>$ due to the statistical uncertainties of the fitted
amplitudes of the various components. There is an additional
uncertainty in the value of $N_{\BB}$, 
evaluated at 1.1\,\%, as well as the fractional uncertainty in the amount
of \BB\ background present (2\,\%). The systematic uncertainties
for the resonance branching fractions ${\cal{B}}_k$ are just the quadratic
sum of the systematic errors for the resonance fit fractions $F_k$ 
and all the contributions to the systematic error for
the total branching fraction ${\cal{B}}^{\pm}_{\rm{tot}}$ except 
the fixed \BB\ background component, since this is already
included in the fit fraction systematics.

For the charge asymmetries, systematic uncertainties from
fit biases, efficiency 
corrections and fluctuations in the background and efficiency histograms
are not included, since they cancel out.
Finally, an uncertainty of 2\,\% is assigned for
the total and fit fraction charge asymmetries due to a
possible detector charge bias, which is determined by finding
the difference between the total number of positively and negatively charged tracks
in the on-resonance data sample.

\section{Summary}
\label{sec:Summary}

The total branching fraction for the charmless decay \BtoPPP\ is measured to be 
$(16.2 \pm 1.2 \pm 0.9) \times 10^{-6}$,
where the first uncertainty is statistical and the second is systematic.
The dominant component in the charmless \BtoPPP\ Dalitz plot is the \rhoI\ resonance.
We have a $3~\sigma$ indication for the presence of the \fzII\ and nonresonant
components. The fit fractions of the resonances \rhoII\ and \fz\ are not
statistically significant.
The decay $B^{\pm} \ra \rhoI \pi^{\pm}$ has a measured branching fraction of
$(8.8 \pm 1.0 \pm 0.6^{+0.1}_{-0.7}) \times 10^{-6}$, which is consistent
with previous measurements~\cite{Belle,Babar1} and theoretical
calculations~\cite{neubert,kramer}. This decay can be used
to help reduce the theoretical uncertainties in the extraction of the CKM angle $\alpha$
from the neutral decays $B^0 \ra \rho^{\pm} \pi^{\mp}$ and 
$B^0 \ra \rho^0 \pi^0$~\cite{SnyderQuin:1993}.
It is found that there is no contribution from the \chiczero\ resonance to 
the \BtoPPP\ Dalitz plot,
which means that the methods advocated in Ref.~\cite{Gronau:1995,Blanco:1998,Blanco:2001}
to measure the CKM angle $\gamma$ are not feasible with our current dataset.
There is also little evidence for contributions from the \fzIII\ and $\sigma$
resonances. Differences in the parameterizations of the \fz\ and nonresonant components
do not significantly affect the results.
Charge asymmetries observed for the total rate and resonance fit
fractions are consistent with zero, 
and 90\,\% confidence-level upper limits are provided for the branching fractions
for resonances that do not have statistically significant fit fractions.
The results presented in this paper supersede those of previous \babar\ analyses.

\section{Acknowledgments}
\label{sec:Acknowledgments}

We are grateful for the 
extraordinary contributions of our \pep2\ colleagues in
achieving the excellent luminosity and machine conditions
that have made this work possible.
The success of this project also relies critically on the 
expertise and dedication of the computing organizations that 
support \babar.
The collaborating institutions wish to thank 
SLAC for its support and the kind hospitality extended to them. 
This work is supported by the
US Department of Energy
and National Science Foundation, the
Natural Sciences and Engineering Research Council (Canada),
Institute of High Energy Physics (China), the
Commissariat \`a l'Energie Atomique and
Institut National de Physique Nucl\'eaire et de Physique des Particules
(France), the
Bundesministerium f\"ur Bildung und Forschung and
Deutsche Forschungsgemeinschaft
(Germany), the
Istituto Nazionale di Fisica Nucleare (Italy),
the Foundation for Fundamental Research on Matter (The Netherlands),
the Research Council of Norway, the
Ministry of Science and Technology of the Russian Federation, and the
Particle Physics and Astronomy Research Council (United Kingdom). 
Individuals have received support from 
CONACyT (Mexico),
the A. P. Sloan Foundation, 
the Research Corporation,
and the Alexander von Humboldt Foundation.



\begin{thebibliography}{99}


\bibitem{CKM:1973}
M.~Kobayashi and T.~Maskawa,
\progtp{49}, 652 (1973).

\bibitem{Gronau:1995}
G.~Eilam, M.~Gronau, R.~R.~Mendel,
\jprl{74}, 4984 (1995).

\bibitem{Blanco:1998}
I.~Bediaga \etal\,
\jprl{81}, 4067 (1998).

\bibitem{Blanco:2001}
R.~E.~Blanco, C.~G\"obel and R.~M\'endez-Galain,
\jprl{86}, 2720 (2001).

\bibitem{SnyderQuin:1993}
A.~E.~Snyder and H.~R.~Quinn,
\jprd{48}, 2139 (1993).

\bibitem{Alde}
GAMS Collaboration, D.~Alde \etal\, 
\plb{397}, 350 (1997).

\bibitem{E791}
E791 Collaboration, E.~M.~Aitala \etal\, 
\jprl{86}, 770 (2001).

\bibitem{BESsigma}
BES Collaboration, M.~Ablikim \etal\,
\plb{598}, 149 (2004).

\bibitem{Belle}
Belle Collaboration, A.~Gordon \etal\,
\plb{542}, 183 (2002).

\bibitem{Babar1}
\babar\ Collaboration, B.~Aubert \etal\,
\jprl{93}, 051802 (2004).

\bibitem{Babar2}
\babar\ Collaboration, B. Aubert \etal\,
\jprl{91}, 051801 (2003).

\bibitem{babardet}
\babar\ Collaboration, B.~Aubert \etal\, 
Nucl.~Instrum.~Meth.~A~{\bf{479}}, 1 (2002).

\bibitem{pep}
PEP-II Conceptual Design Report, SLAC-R-418 (1993).

\bibitem{chargeConjugate}
Charge conjugates are included implicitly for the calibration and charm veto modes.



\bibitem{FoxWolfram}
G.~C.~Fox and S.~Wolfram, \jprl{41}, 1581 (1978).


\bibitem{fisher}
R.~A.~Fisher,
Annals Eugen. {\bf 7}, 179 (1936);\\
G.~Cowan,
\emph{Statistical Data Analysis}, (Oxford University Press, 1998), p. 51.

\bibitem{tagging}
\babar\ Collaboration, B.~Aubert \etal\,
\jprl{~89}, 201802 (2002).

\bibitem{pdg2004}
Particle Data Group, S.~Eidelman \etal\,
\plb{592}, 1 (2004).


\bibitem{argus}
ARGUS Collaboration, H.~Albrecht \etal\,
\zp{C~48}, 543 (1990).


\bibitem{Dalitz}
R.~H.~Dalitz, Phil.~Mag.~{\bf 44}, 1068 (1953).

\bibitem{blatt}
J.~M.~Blatt and V.~F.~Weisskopf, \emph{Theoretical Nuclear Physics},
J.~Wiley \& Sons, New York (1952).

\bibitem{flatte}
S.~M.~Flatt\'e, 
\plb{63}, 224 (1976).

\bibitem{Zemach}
C.~Zemach, \pr{133}, B1201 (1964).

\bibitem{Zemachpdg}
D.~Asner (2003), hep-ex/0410014.


\bibitem{Colangelo}
G.~Colangelo \etal\, 
\npb{603}, 125 (2001).

\bibitem{bugg}
D.~V.~Bugg,
\plb{572}, 1 (2003).

\bibitem{BBbarPairs}
We assume that the $\Upsilon(4S)$ decays equally
to neutral and charged $B$ meson pairs.\\
\babar\ Collaboration, B.~Aubert \etal\,
\jprd{69}, 071101 (2004).

\bibitem{neubert}
M.~Beneke and M.~Neubert,
\npb{675}, 333 (2003).

\bibitem{kramer}
G.~Kramer and C-D.~L\"u,
\ijmpa{13}, 3361 (1998).


\bibitem{WA76}
WA76 Collaboration, T.~A.~Armstrong \etal\,
\zpc{51}, 351 (1991).

\bibitem{E791Flatte}
E791 Collaboration, E.~M~Aitala \etal\,
\jprl{86}, 765 (2001).

\bibitem{BelleNR}
Belle Collaboration, A.~Garmash \etal\,
\jprd{71}, 092003 (2005).


\end{thebibliography}
\end{document}